\documentclass{article}

\usepackage{amsmath}
\usepackage{amssymb}
\usepackage{PRIMEarxiv}
 \usepackage{setspace}
\usepackage[utf8]{inputenc} % allow utf-8 input
\usepackage[T1]{fontenc}    % use 8-bit T1 fonts
\usepackage{hyperref}       % hyperlinks
\usepackage{url}            % simple URL typesetting
\usepackage{booktabs} % professional-quality tables
\usepackage{amsfonts}       % blackboard math symbols
\usepackage{tabularx}       %allows table width to be adjusted
\usepackage{nicefrac}       % compact symbols for 1/2, etc.
\usepackage{caption}
\usepackage{microtype}      % microtypography
\usepackage{lipsum}
\usepackage{fancyhdr}       % header
\usepackage{graphicx}       % graphics
\usepackage{float}

\graphicspath{{media/}}     % organize your images and other figures under media/ folder

\title{Unraveling Factors Influencing Shooting Incidents: Preliminary Analysis and Insights
\thanks{\textit{\underline{Citation}}: 
\textbf{Hernandez, Carothers. Unraveling Factors Influencing Shooting Incidents: Preliminary Analysis and Insights}} 
}

\author{
  Hernandez Jr, Robert and Carothers, Linn \\
  Department Of Applied Mathematics \\
  California Baptist University \\
  Riverside, CA\\
  \\
}
\begin{document}
\maketitle

\begin{abstract}
The following is a write up of the progress of modeling data from the K12 organization \cite{Riedman_2023}. Data was characterized and investigated for statistically significant factors. The incident data was spilt into three sets: the entire set of incidents, incidents from 1966 - 2017, and incidents from 2018 - 2023. This was done in an attempt to discern key factors for the acceleration of incidents over the last several years. The data set was cleaned and processed primarily through RStudio. The individual factors were studied and subjected to statistical analysis where appropriate. As it turns out, there are differences between media portrayals of shooters and actual shooters. Then, multiple regression techniques were performed then followed by ANOVA of the models to determine statistically significant independent variables and their influence on casualties. Thus far, linear regression and negative binomial regression have been attempted. Further refining of the methods will be necessary for Poisson regression and logistic regression to be viably attempted. At this point in time a common theme among each of the models is the presence of targeted attacks affecting casualties. Further study can lead to improved safe guarding strategies to eliminate or minimize casualties. Further, increased understanding of shooter demographics can also lead to outreach and prevention programs. 

\end{abstract}

% keywords can be removed

\section{Introduction}
\paragraph{}In the United States of America, gun violence has continued to be a leading cause of preventable death\cite{InjuryFacts2023}. The fatalities are made up of suicides, homicides, and gang violence, among others. Of all occurrences of gun violence, mass shootings have seen steep growth over the past several years\cite{fivethirtyeight_massshootings}. A subset of mass shootings is defined by those that take place at schools. School shootings, while not a new phenomenon, have been steadily and steeply increasing over the last several years \cite{fivethirtyeight_massshootings}. 
\paragraph{}School shootings carry with them a unique set of fears and anxieties. Millions of children funnel into schools every day. Parents and guardians entrust the safety of their children to the schools, but have however become less confident that their schools are a safe place to be\cite{ADAA2022}. With the rise of active shooter events in schools, active shooter drills have become an increasing practice on par with drills for natural disasters. With this increased exposure to the concept of a person or group of people opening fire on school grounds, there is increased anxiety\cite{ADAA2022} among school-aged children that they are not in a safe place and that it may be only a matter of time before they become victims of a school shooting.  Many questions about what can be done arise for how these events can be prevented and casualties limited. 
\paragraph{}A potential factor in decreasing the occurrence and severity of a school shooting is to explore what the data from past events can tell us. There are many databases that hold information collected from past events of mass and school shootings that can be analyzed for patterns and commonalities. Mathematical modeling using computer software packages to tease out contributing factors is a growing area of study. In analyzing this data, there are many questions to answer. Which model is the most accurate? Which variables carry the most significance in predicting the outcome of a school shooting? How can the findings be applied to prevention measures at every level of government? What, if any, explanations are there for the dramatic increase in these events over the last several years? This paper seeks to explore some of these questions with a focus on a single dataset of school shootings that encompass over 6 decades of details about school shootings. Multiple computational regression techniques and statistical analysis tools such as ANOVA (analysis of variance) were used with the intent to uncover which ways are more effective in predicting the severity and occurrence of a school shooting. 
\paragraph{}The K12 School Shooting database\cite{Riedman_2023} details over 2500 separate events of school shootings over the past 6 decades. This dataset is made up of dozens of potential variables and countless potential combinations of these variables for statistical analysis. Using established techniques with R software (version 4.2.3 and 4.3.3), this paper attempts to contribute to the narrowing of what these variables are and what they can tell us.

\section{Methods}
To study the mathematical relationships between various outcomes and potential influencing variables, a large dataset was requested from the K-12 School Shooting database via email request. Once the dataset was obtained, the data was prepared for analysis and modeling. The initial dataset consisted of over 50 variables and nearly 2,600 individual events across 4 Microsoft Excel spreadsheets. The methods for preparing the data varied depending on the variable in question. The final dataset was a merged into a  .csv formatted file that could be loaded directly into RStudio for manipulation and analysis. Care was taken to omit duplicate incidents and merge data when appropriate to a single incident that would retain the information needed for analysis. Individual aspects of the dataset depending on the variable in use were studied for statistical characteristics when the variable was numerical and further investigated for statistical significance, when the variable was categorical in format, proportions were calculated to get a grasp on the scale of differences between the categorical groups, so that any trends and possible characteristic consistencies could be noted. Later, the dataset a split between events that took place up until the year 2017 and events that took place from 2018 to present, so that potential trends and changes in the pattern of events, especially in light of the marked increase in school shooting events from 2018. The regression models were generated using each of the 3 subsets of data (through 2017, 2018 to present, and all events), then ANOVA was performed on each to investigate for impacting variables. Finally, the linear regression model for all events was further dissected among the most significant variables to test for actual influence on casualty outcomes. 
\label{sec:headings}

\subsection{Characteristics of The Dataset}

\begin{figure}[H]
  \centering
  \noindent % Prevents indentation for the first line
  \begin{minipage}{.5\textwidth} % Begin the minipage for the first image
    \centering
    \includegraphics[width=.9\linewidth]{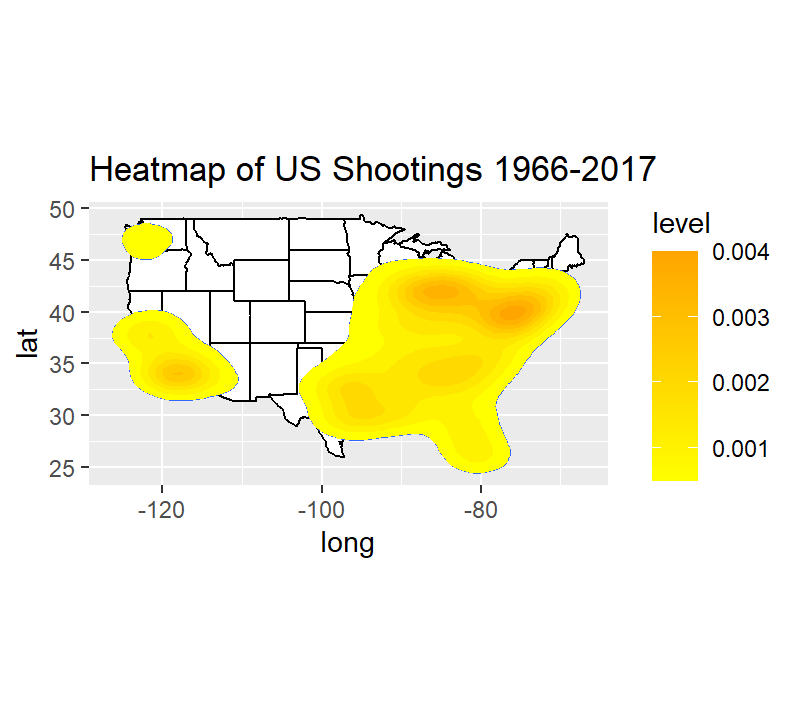}
    % Image-specific description can still be added here if needed
  \end{minipage}% This comment prevents adding space between the minipages
  \begin{minipage}{.5\textwidth} % Begin the minipage for the second image
    \centering
    \includegraphics[width=.9\linewidth]{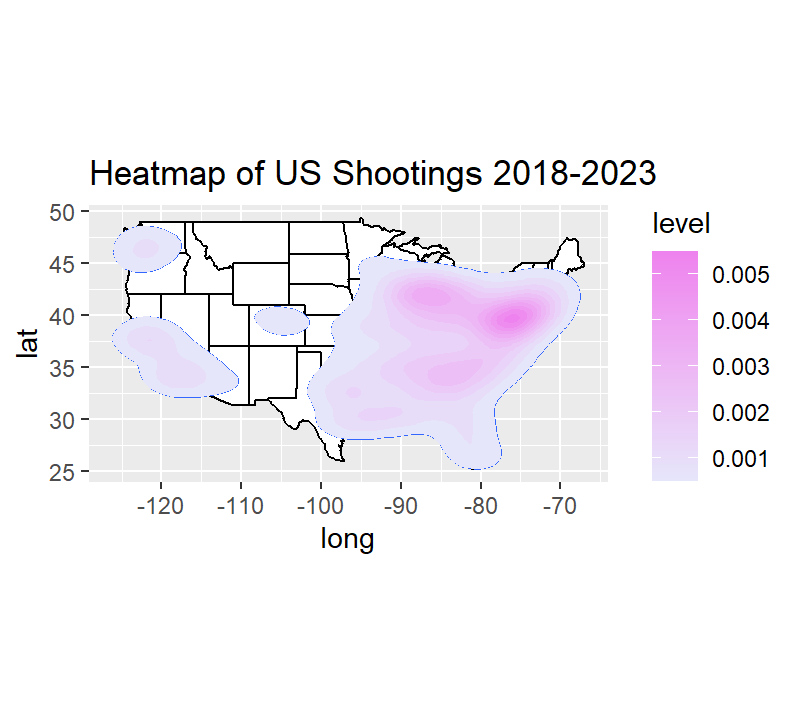}
    % Image-specific description can still be added here if needed
  \end{minipage}
  \caption{Heat Maps of Events Broken up by Time Period} % This adds the caption below both images
  \label{fig:heatmapsoldandnew} 
\end{figure}

\begin{figure}[H] % 'ht' tells LaTeX to place the figure here if possible, at the top of the page otherwise
  \centering
  \includegraphics[width=.6\linewidth]{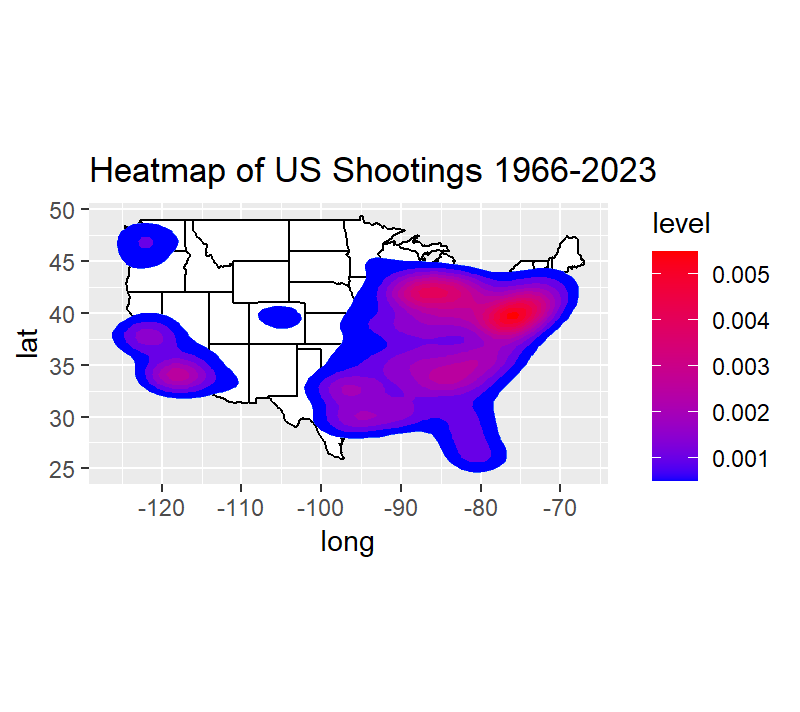} % Adjust the width as necessary
  \caption{}
  \label{fig:AllHeatMap}
\end{figure}

\paragraph{Geographical Heat Maps}
While not used for the analysis in this study, it would be useful to note the prevalence of certain regions of the United States that have higher occurrences of school shootings. The database\cite{Riedman_2023} included geographical coordinates, for each event in the set. A geographical heat map\ref{fig:AllHeatMap} was generated in RStudio which highlights some of the zones in the United States that have higher occurrences of school shootings. Further study should be performed and refined to map the prevalence of events based upon weapons of choice, firearms laws, demographic changes over time. The growth in event frequency could also be studied for regional trends. A second set of heat maps\ref{fig:heatmapsoldandnew} was generated to look for differences. One noticeable difference is that there is an increase in the region of Colorado in the latter events from 2018-2023. However, aside form this, the regional changes are minimal and seem to be consistent from one time period to another. It would be worth investigating what differences led to the uptick in events in the Colorado region.

% Center the third image below the first two

\begin{figure}[H]
    \centering
    \includegraphics[width=0.40\textwidth]{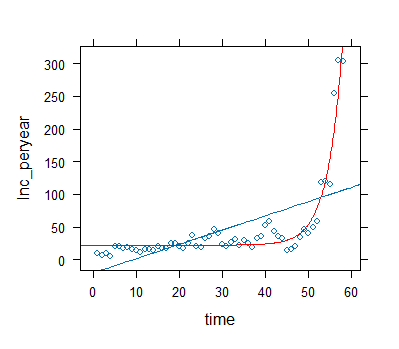}
    \caption{Scatter plot of incidents per year with linear and exponential curve fit}
    \label{eventfrequency}
    
\end{figure}

\paragraph{Modeling of Shooting Frequency Data}
Using fitModel() function in R, (included in Mosaic Package) incidents per year were modeled linearly and exponentially. The exponential model yielded the best fitting curve\ref{eventfrequency}. The increase in event frequency is stark beginning in the year 2018 or so. This can be almost characterized as an inflection point\cite{katsiyannis2023examination} in the data considering the relative stability in the shooting events in the years leading up to this time period. 

Linear Equation: $y=2.1622t-19.3702$ \\ Exponential Equation: $y=22.49+0.00001855e^{0.2872t}$

\paragraph{Incidents Per Month}
The dataset includes a column of the date that each incident occurred. In an analysis of the number of incidents that occurred per month \ref{monthlydistribution}, July is the lowest, with 70 and September is the highest, with 344, and mean, median, and standard deviation 214.67, 217, and 79.59, respectively. 

\begin{table}[H]
\centering
\caption{Incidents Per month}
\label{monthlydistribution}
\begin{tabularx}{\textwidth}{XrXr}
\toprule
\textbf{Month} & \textbf{Occurrences} & \textbf{Month} & \textbf{Occurrences} \\
\midrule
January & 266 & July & 70 \\
February & 231 & August & 172 \\
March & 235 & September & 344 \\
April & 201 & October & 319 \\
May & 253 & November & 203 \\
June & 100 & December & 182 \\
\addlinespace
\multicolumn{4}{c}{\textbf{Summary Statistics}} \\
\midrule
\multicolumn{2}{c}{Mean: 214.67} & \multicolumn{2}{c}{Median: 217.00} \\
\multicolumn{4}{c}{Standard Deviation: 79.59} \\
\bottomrule
\end{tabularx}
\end{table}

A  $\chi^{2}$ test was performed to see if there was a statistical significance in the number of shootings that take place in a given month. This information can imply seasonal trends,especially since this data is focused on the United States. The table\ref{table:significant_months} shows that January, May, September, and October have a higher than expected rate of shootings while June, July, August, and December have lower than expected shooting incidents. This implies a some hot periods throughout the year from mid-winter, late spring, fall months. Most schools in the United States are either closed or under reduced operations during the Summer months, which could explain why there would be a lower than expected incidence rate for June -  August. 

\begin{table}[H]
\centering
\caption{Statistical Significance of Monthly Incident Counts $\chi^{2}$ = 324.62, df = 11, p-value < 2.2e-16}
\label{table:significant_months}
\begin{tabular}{lrrrl}
\toprule
\textbf{Month} & \textbf{Observed} & \textbf{Expected} & \textbf{Adj. Residual} & \textbf{Higher/Lower} \\
\midrule
January   & 266 & 214.6667 & 3.50 & Higher \\
May       & 253 & 214.6667 & 2.62 & Higher \\
June      & 100 & 214.6667 & -7.83 & Lower \\
July      & 70  & 214.6667 & -9.87 & Lower \\
August    & 172 & 214.6667 & -2.91 & Lower \\
September & 344 & 214.6667 & 8.83 & Higher \\
October   & 319 & 214.6667 & 7.12 & Higher \\
December  & 182 & 214.6667 & -2.23 & Lower \\
\bottomrule
\end{tabular}
\end{table}

\paragraph{Casualties}
The dataset includes information about the number of people killed or wounded in each incident. There are a total of 780 killed, and 2152 wounded. Below\ref{tab:victims_summary} is a table of statistical data about the number of killed and wounded. While many incidents that did not result in casualties, there are outlier values. There were 5 incidents of double-digit fatalities, the highest of which was 26. For the wounded, there were 16 double-digit wounded with the highest being 74 followed by 30 for the next highest. A breakdown\ref{tab:VitimDemo} of the demographic and outcome information of the victims involved shows that of those affected in all of the events 60.28 percent were wounded, while 22.18 percent were fatally wounded. In 14 percent of the dataset there is no specific data available. The victims skew male with over 3/4 of the total. Student's make up 2/3 of the victims while teachers make up 5.47 percent. Over half of the victims are teens which is also a significant category for the shooters, as well. Finally, black victims make up the majority followed by white with 50.29 and 35.96 percent, respectively.

\begin{table}[H]
  \centering
  \caption{Summary Statistics of Victims}
  \begin{tabular}{@{}llll@{}}
    \toprule
    \textbf{Statistic} & \textbf{Killed} & \textbf{Wounded} \\ 
    \midrule
    Mean & 0.3028 & 0.8354 \\
    Median & 0 & 0 \\
    Maximum & 26 & 74 \\
    Standard Deviation & 0.9742 & 2.1831 \\
    \bottomrule
  \end{tabular}
  \label{tab:victims_summary}
\end{table}

\begin{table}[H]
\centering
\begin{tabular}{|l|r||l|r||l|r||l|r||l|r|}
\hline
\multicolumn{2}{|c||}{Injury} & \multicolumn{2}{c||}{Gender} & \multicolumn{2}{c||}{School Affiliation} & \multicolumn{2}{c||}{Age} & \multicolumn{2}{c|}{Race} \\
\hline
Category & \% & Category & \% & Category & \% & Category & \% & Category & \% \\
\hline
Wounded & 60.28 & Male & 75.94 & Student & 66.30 & Teen & 54.02 & Black & 50.29 \\
Fatal & 22.18 & Female & 22.39 & No Relation & 11.85 & Adult & 33.71 & White & 35.96 \\
None & 14.00 &  &  & Teacher & 5.47 & Child & 12.18 & Hispanic & 10.53 \\
Non-gunshot & 3.54 &  &  & Nonstudent & 4.24 &  &  & Asian & 2.63 \\
 &  &  &  & Other Staff & 2.48 &  &  & Unknown & 0.29 \\
 &  &  &  &  &  &  &  & Latino & 0.29 \\
\hline
\end{tabular}
\caption{Breakdown of Demographic Information of Shooting Victims}
\label{tab:VitimDemo}
\end{table}

\paragraph{Shooter Demographics}
An important aspect of studying the school shooting data is a breakdown and analysis of the shooter(s) themselves. While the dataset contains information on the shooter's age, race, gender, gang affiliation, student status, and the result of the incident, there are large gaps within the set. These will be addressed for each category discussed in this paper.

\begin{figure}[H]
\begin{minipage}{0.45\textwidth}
\centering
\includegraphics[width=\linewidth]{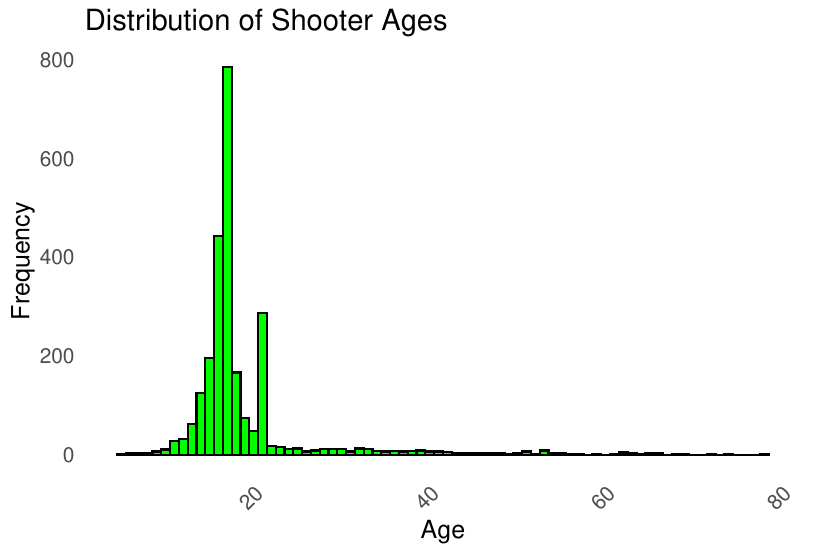}
\caption{Histogram of Shooter Ages}
\label{fig:shooteragedist}
\end{minipage}
\hfill
\begin{minipage}{0.45\textwidth}
\centering
\begin{tabular}{lc}
\hline
\textbf{Test} & \textbf{Result} \\
\hline
Shapiro-Wilk Test Stat & 0.66 \\
P-value & \(< 0.0001\) \\
\hline
\multicolumn{2}{p{7cm}}{\textit{Note: A p-value less than 0.05 typically indicates that the data are not normally distributed.}} \\
\end{tabular}
\caption{Normality Test Results for Shooter Ages}
\label{tab:normality_test_shooter_ages}
\end{minipage}
\end{figure}

\paragraph{Shooter's Age} The ages of the shooters predominantly late teens. The median age is 17 years old and mean age is 20.8 years old with a standard deviation of 10.03 years. Of the non-numerical values, there are entries that contain labels (teen, adult etc.) these were taken into account later in the dataset before modeling was attempted. The histogram of shooter age\ref{fig:shooteragedist} shows an approximate normal distribution. The spike in the 21 year old age category can be attributed to the age being assigned to all the incidents which are labeled "Adult." However, after completing a Shapiro-Wilk\cite{shapiro1965analysis} test\ref{tab:normality_test_shooter_ages} on the data to check for normal distribution, the ages are not appear to be normally distributed. 

\begin{table}[H]
\centering
\begin{tabular}{lr}
\hline
\textbf{Statistic} & \textbf{Value} \\
\hline
Mean Age & 20.08 years \\
Median Age & 17.0 years \\
Mode Age & 17.0 years \\
Standard Deviation & 10.03 years \\
NA Count & 1272 \\
\hline
\end{tabular}
\caption{Summary Statistics for Age}
\label{tab:age_summary}
\end{table}

\paragraph{Shooter's Gender} The shooter's gender was predominately male, however other genders are represented in the data set. 

\paragraph{Shooter's Race}
Unfortunately, there is a dearth of information regarding the race of the shooters. Over two-thirds of the shooters race is unknown in the dataset. The breakdown of the available race information is included in a data table \ref{tab:race_breakdown_known}. Furthermore, it is noted that there is a higher proportion of black shooters in the dataset. This has actually been proven to be a more recent phenomenon\cite{katsiyannis2023examination} which is also representative of the victims as well. 

\begin{table}[H]
\centering
\begin{minipage}{0.49\textwidth}
\centering
\begin{tabular}{lc}
\hline
\textbf{Gender} & \textbf{Count} \\
\hline
Male & 2199 \\
Female & 97 \\
Transgender & 2 \\
Unknown & 648 \\
\hline
\end{tabular}
\caption{Shooter Gender Breakdown}
\label{tab:gender_breakdown}
\end{minipage}
\hfill
\begin{minipage}{0.49\textwidth}
\centering
\begin{tabular}{lcc}
\hline
\textbf{Race} & \textbf{Count} & \textbf{Percentage (\%)} \\
\hline
Black & 369 & 50.55 \\
White & 247 & 33.84 \\
Hispanic & 87 & 11.92 \\
Asian & 14 & 1.92 \\
Other & 6 & 0.82 \\
Indigenous & 4 & 0.55 \\
Hawaiian/Pacific Island & 2 & 0.27 \\
Middle Eastern & 1 & 0.14 \\
\hline
\end{tabular}
\caption{Shooter Race Breakdown (Excluding Unknown)}
\label{tab:race_breakdown_known}
\end{minipage}
\end{table}

\paragraph{Shooter's Affiliation} The dataset included,where known, the affiliations of shooters\ref{table:school_affiliation} to schools in which a shooting event took place. There were several forms of relationships that could potentially provide insight into the context of and individual event or expose patterns. The majority of shooters were current students (1,106), which demonstrates how many of these attacks come from within the school. The category 'No Relation', meaning not affiliated with the schools had 575 of the shooters which also shows that outside threats are a real concern. 'Nonstudent Using Athletic Facilities/Attending Game' (130 instances), categories are also a non-trivial portion of the events which means that anytime there is a access to the school there is risk involved. 
\begin{table}[H]
\centering
\begin{tabular}{lrlr}
\hline
\multicolumn{2}{c}{\textbf{School Affiliation Breakdown}} & \multicolumn{2}{c}{\textbf{Instances}} \\
\hline
Student & 1106 & Nonstudent Using Athletic Facilities/Attending Game & 130 \\
No Relation & 575 & Former Student & 97 \\
Unknown & 331 & Parent & 62 \\
Nonstudent & 36 & Police Officer/SRO & 53 \\
Rival School Student & 30 & Intimate Relationship & 38 \\
Relative & 28 & Other Staff & 26 \\
Teacher & 19 & Other Student & 11 \\
Gang Member & 7 & Security Guard & 6 \\
Former Teacher & 2 & Principal/Vice-Principal & 1 \\
Friend & 1 & Hitman & 1 \\
\multicolumn{2}{l}{\textit{NaN} (Missing Values)} & \multicolumn{2}{r}{388} \\
\hline
\end{tabular}
\caption{Breakdown of Shooter School Affiliation}
\label{table:school_affiliation}
\end{table}

\paragraph{Additional Shooter Background}
According to the data set, the shooter backgrounds skew high for non-bullied, non-gang-related, and those without a history of domestic violence. These are all commonly cited factors in popular culture that the data does not bear out. Findings such as this demonstrate a need to increased understanding in the individuals involved and at risk for participating in school violence.

\begin{table}[H]
\centering
\begin{tabular}{lcc}
\hline
Category & Yes (\%) & No (\%) \\ 
\hline
Bullied & 4.55148 & 95.44852 \\
Domestic\_Violence & 5.226904 & 94.773096 \\
Gang\_Related & 13.558442 & 86.441558 \\
\hline
\end{tabular}
\caption{Additional Shooter Background Information}
\label{tab:ShooterHist}
\end{table}

\paragraph{Shooting Event Details and Outcomes}
The data set can also be used to construct a "typical" school shooting based on the predominant characteristics. For instance, a significant majority (61.63\%) of shootings occurred during class sessions\ref{tab:EventsDeatailsTimeandStyle}, this highlights the vulnerability of students when they are gathered in the classroom on on school grounds, in general. Shooter's primarily act alone according to the data set as 77.59\% of events involve single shooters. Hostages\ref{tab:EventsDeatailsTimeandStyle} are rarely taken and account for only 2.36\% of events. Shootings where the shooter was killed\ref{tab:EventsDeatailsTimeandStyle} constituted a small proportion (9.86\%) meaning the gunman either flees unharmed or is neutralized non-lethally. Handguns\ref{tab:weapon_type} appear to be the main choice by shooter as they were used in 72.65\% of incidents, while instances involving multiple firearms had much lower occurrences. There was also a notable proportion of incidents (9.43\%) where the weapon type was not recorded/unknown. Over half of the incidents (56.77\%) involved targeted\ref{tab:targets} victims, indicating premeditated and/or personal motivations for attacking, while random shootings accounted for only 16.30\%. The majority of shootings (60.67\%)\ref{tab:location_type} occurred outside but still on school grounds, with nearly 1/3 inside a school building (29.99\%). Considering the selection of casualties as the dependent variable for regression modeling, the statistical significance of these situational data points is of high interest.

\begin{table}[H]
\centering
\begin{tabular}{lcc}
\hline
Category & Yes (\%) & No (\%) \\ 
\hline
During\_Classes & 61.632342 & 38.367658 \\
Accomplice & 22.408330 & 77.591670 \\
Hostages & 2.358311 & 97.641689 \\
Shooter\_Killed & 9.856495 & 90.143505 \\
\hline
\end{tabular}
\caption{Shooting Event Details }
\label{tab:EventsDeatailsTimeandStyle}
\end{table}

\begin{table}[H]
\centering
\begin{minipage}[b]{0.32\linewidth} % The minipage environment for the first table
\centering
\begin{tabular}{lc}
\toprule
Weapon Type & Percentage \\
\midrule
handgun & 72.65 \\
multiple handguns & 2.30 \\
multiple rifles & 0.37 \\
multiple unknown & 1.19 \\
no data & 9.43 \\
other & 4.92 \\
rifle & 5.21 \\
shotgun & 2.95 \\
unknown & 0.98 \\
\bottomrule
\end{tabular}
\caption{Weapon Type}
\label{tab:weapon_type}
\end{minipage}
\hfill
\begin{minipage}[b]{0.32\linewidth} % The minipage environment for the second table
\centering
\begin{tabular}{lc}
\toprule
Targets & Percentage \\
\midrule
both & 14.38 \\
neither & 12.55 \\
random shooting & 16.30 \\
victims targeted & 56.77 \\
\bottomrule
\end{tabular}
\caption{Targets}
\label{tab:targets}
\end{minipage}
\hfill
\begin{minipage}[b]{0.32\linewidth} % The minipage environment for the third table
\centering
\begin{tabular}{lc}
\toprule
Location Type & Percentage \\
\midrule
Both Inside/Outside & 1.06 \\
Inside & 29.99 \\
Off Campus & 2.84 \\
Outside & 60.67 \\
School Bus & 5.18 \\
\bottomrule
\end{tabular}
\caption{Location Type}
\label{tab:location_type}
\end{minipage}
\end{table}

\subsection{Preparing Data-set for Modeling}
\paragraph{The Dataset}
The dataset studied was the K-12 School Shooting Database provided by an email request to David Reidman.  The dataset arrived in the form of an Microsoft Excel spreadsheet. The data was separated into 4 spreadsheets containing details of 2584 school shooting incidents dating back to 1966, each with a unique identifying code. The incident sheet contained available details for each incident including narrative details, when available, along with data on the number for victims, the outcome of the shooting, the duration, and number of shots fired. The shooter sheet contained demographic information and situational details of the shooters involved in each incident. There were many incidents that involved more than one shooter. The weapons tab contained any available details about the types of firearms used in the incident. The victims tab contained demographic information for any available data about the victims in each incident. 

\paragraph{Shooter Information}
The demographic information for each shooter was linked to the unique incident ID. In order to make the data usable for linear modeling, the data needed to be “cleaned up” to account for missing or incomplete details. Where multiple shooters were involved, demographic information was merged using multiple techniques.

\paragraph{Shooter Age}
For the shooter’s age column, cells either contained a number, when known, a description (child, teen, adult), or the cell was blank. To streamline this column, first the statistical information was collected on the datapoints that contained numerical information. The mean, median, and standard deviation was calculated. Then, for blank cells, the median age of 17 was used. For cells that contained descriptive information, the data was split into demographics and new statistics were calculated. A cell containing “child” was changed to the median age of the datapoints that contained shooters under 12 years of age. For cells containing “teen” the median age was used for shooters between the age of 13 and 17. For cells containing “adult” the median age for shooters over the age of 18 was used. Finally, the average age of the shooters involve was used when there were more than one in an incident.
\paragraph{Shooter Gender}The shooter’s gender column was sorted and combined for events that involved multiple shooters. For example, and occurrence of all male perpetrators would be labeled as “male” and all female perpetrators “female.” Where there were a combination of genders (of which there were few) they were labeled “multiple.” Where no data was present, “unknown” was used. 
\paragraph{Shooter Race} The race of the shooter, when known, was included in the dataset for each event. When multiple shooters of the same race were involved, they were labeled as said race. In an event where there were a combination of races among the shooters, they were labeled as “multiple.” If it was unknown, the event was labeled “NA” which is handled by R in multiple ways, depending on the technique being used. 
\paragraph{Yes/No Variables}There were several variables in the dataset that contained binary data. Either yes or no where available. These included whether or not the event took place while classes were in session, if a police officer involved, if the perpetrator had a history of bullying or domestic violence, if there was an accomplice, or if the incident was tied to gang violence. In all of these instances, if no data was available, the event was labeled “NA” which is handled by R in different ways depending on the model being used. 
\paragraph{Weapon Type}For the weapon type variable. The weapons were labeled by class (e.g. handgun, shotgun). Where multiple weapons where involved, the event was labeled “multiple” along with the class of weapon. This applies to unknown event data as well, whether there was one or multiple weapons used.

\subsection{Modeling and ANOVA}
In this paper, there are 3 main goals for the data modeling, 1) determining which modeling technique is the best at fitting the data for selected dependent variables, 2) determining which independent variables had the most statistical impact, and 3) after splitting the data between the years prior to 2018, and 2018 to present, which variables could be used to explain the significant increase school shooting events. 
\subsubsection{Dependent Variable Selection}
The K12 School Shooting dataset contains over 2500 individual events with over 50 potential independent variables, depending on which relationships are being examined. It was decided that the first relationship that should be examined was the resultant casualties from the incidents. This would be the dependent variable that would be used for the modeling equations. Secondly, the number of incidents per year would be modeled over time by multiple modeling techniques to analyze for any trends and potentially predictive for future incident counts.  

\subsubsection{The Models}
With the dataset cleaned and coded. The data was run through multiple modeling functions in RStudio. As this is an ongoing study, the models currently listed are not all inclusive. The models used were:
\begin{itemize}
    \item Linear (least squares)
    \item Negative Binomial
\end{itemize}
After each model was processed. It was analyzed using an ANOVA (Analysis of Variance) function and other techniques specific to the model (discussed in results section) to determine the validity of the model, the statistical significance of the independent variables, and iteratively remove independent variables that were not significant from the model.

\subsubsection{Linear Modeling}

The linear model (as implemented in the \texttt{lm()} function in R) is given by the equation:
\begin{equation}
    y = \beta_0 + \beta_1 x_1 + \beta_2 x_2 + \ldots + \beta_p x_p + \epsilon
\end{equation}
where:
\begin{itemize}
    \item $y$ is the dependent variable.
    \item $\beta_0$ is the intercept term.
    \item $\beta_1, \beta_2, \ldots, \beta_p$ are the coefficients of the independent variables $x_1, x_2, \ldots, x_p$.
    \item $\epsilon$ is the error term.
\end{itemize}
The coefficients $\beta$ are estimated using the least squares method.

\subsubsection{Negative Binomial Regression}
The negative binomial distribution for a random variable \(Y\) is given by the equation:
\begin{equation}
    P(Y = y) = \binom{y + r - 1}{y} (1-p)^r p^y
\end{equation}
where \( y \) is the count, \( r \) is the number of failures until the experiment is stopped, and \( p \) is the probability of success. In this case, that means the casualties.

The logarithmic link function in negative binomial regression relates the expected value of \(Y\) to the predictors:
\begin{equation}
    \log(\mu) = X\beta
\end{equation}
or equivalently,
\begin{equation}
    \mu = e^{X\beta}
\end{equation}
where \( \mu \) is the expected value of \(Y\), \( X \) is the matrix of predictors, and \( \beta \) are the regression coefficients.

The variance of \(Y\) in the negative binomial model is adjusted for over-dispersion:
\begin{equation}
    Var(Y) = \mu + \alpha \mu^2
\end{equation}
where \( \alpha \) is the over-dispersion parameter.

\section{Results}
 
\subsection{Linear Regression}
The data was analyzed in RStudio with a linear model for all events, events from 1966 - 2017, and 2018 - 2023. After the model was generated. ANOVA was performed with the methods outlined by Farraway \cite{faraw2015practical}. Then, variance inflation factors were examined. Also correlation factors were also inspected to check for variables that may have strong correlations to each other.

\subsubsection{All Events in Dataset}
\begin{figure}[H]
    \centering
    % Adjust widths as needed to fit side by side, e.g., 0.45\textwidth
    \includegraphics[width=0.35\textwidth]{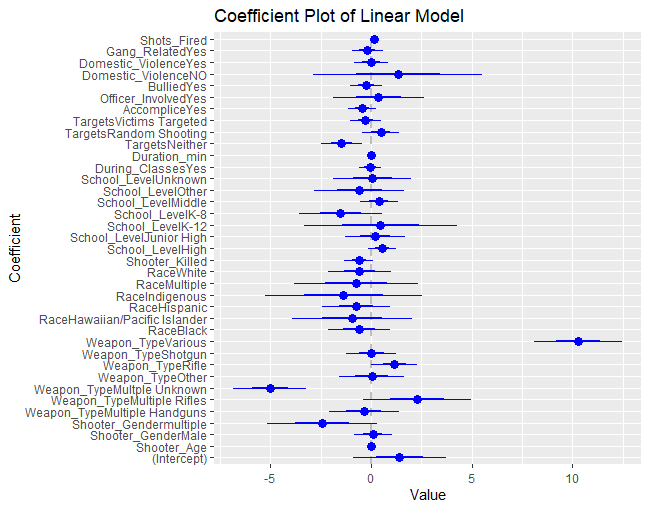}
    \caption{} % Add your caption text here
    \label{COEFLIN}
\end{figure}

After modeling the data for all events linearly, the values that carried the highest likelihood of significance for casualties in a shooting event were:
\begin{itemize}
    \item Shooter Gender
    \item Weapon Type
    \item Specific Targeting of victims
    \item History of Bullying
    \item Number of shots fired
    \item Presence of an accomplice
\end{itemize}
The coefficient matrix \ref{COEFLIN} displays point plots of estimated coefficients with error bars. Lines that cross over the zero line are considered not statistically impactful. Points to the left show a negative correlation and points to the right show a positive correlation. The ANOVA table\ref{AllEventsANOVA} shows that 3 of the factors have a higher statistical significance (Weapon Type, Targets, and Shots fired) on the number of casualties in a given event.

\begin{table}[H]
\centering
\caption{ANOVA Table for All Events}
\label{AllEventsANOVA}
\begin{tabular}{|l|c|c|c|c|c|}
\hline
\textbf{Variable} & \textbf{Df} & \textbf{Sum Sq} & \textbf{Mean Sq} & \textbf{F value} & \textbf{Pr(>F)} \\ 
\hline
Shooter\_Gender & 3 & 13.47 & 4.49 & 2.6289 & 0.049288 * \\ 
\hline
Weapon\_Type & 7 & 1639.10 & 234.16 & 137.0833 & < 2.2e-16 *** \\ 
\hline
Targets & 3 & 233.61 & 77.87 & 45.5870 & < 2.2e-16 *** \\ 
\hline
Accomplice & 2 & 6.72 & 3.36 & 1.9667 & 0.140725 \\ 
\hline
Bullied & 1 & 16.83 & 16.83 & 9.8530 & 0.001769 ** \\ 
\hline
Shots\_Fired & 1 & 594.01 & 594.01 & 347.7551 & < 2.2e-16 *** \\ 
\hline
Residuals & 676 & 1154.70 & 1.71 & & \\ 
\hline
\end{tabular}
\end{table}

However, the factors that showed the most significance (as denoted by the "***" beside the p-value column in the ANOVA table \ref{AllEventsANOVA}. So, the variables Targets, Weapon type, and Shots Fired were then subjected to a 3 factor ANOVA\cite{faraw2015practical}. The results \ref{3FactorAll} further showed a relationship between these factors and casualties in an event.

\begin{table}[H]
\centering
\begin{tabular}{lrrrrr}
\hline
Factor                             & Df   & Sum Sq  & Mean Sq & F value & Pr(>F)        \\ \hline
Targets                            & 3    & 699.0   & 233.0   & 101.297 & $< 2e-16$ *** \\
Weapon\_Type                        & 7    & 1578.1  & 225.4   & 98.011  & $< 2e-16$ *** \\
Shots\_Fired                        & 1    & 1802.0  & 1802.0  & 783.389 & $< 2e-16$ *** \\
Targets:Weapon\_Type                & 20   & 187.3   & 9.4     & 4.072   & $5.00e-09$ ***\\
Targets:Shots\_Fired                & 3    & 119.7   & 39.9    & 17.342  & $5.01e-11$ ***\\
Weapon\_Type:Shots\_Fired            & 7    & 752.9   & 107.6   & 46.762  & $< 2e-16$ *** \\
Targets:Weapon\_Type:Shots\_Fired    & 14   & 188.5   & 13.5    & 5.853   & $3.26e-11$ ***\\
Residuals                          & 1201 & 2762.6  & 2.3     &         &               \\ \hline
\end{tabular}
\caption{3-factor ANOVA Results}
\label{3FactorAll}
\end{table}

A Welch's t-test \cite{ruxton2006unequal} was performed between whether or not victims were targeted in events. The results show there is no statistical significance in casualty outcomes in a targeted attack. Fail to Reject the null hypothesis that a targeted attack nets more casualties.

\begin{table}[H]
\centering
\begin{tabular}{|l|l|}
\hline
\textbf{Metric}                      & \textbf{Value}           \\ \hline
t-value                              & -0.6226                  \\ \hline
Degrees of freedom (df)              & 678.48                   \\ \hline
p-value                              & 0.5338                   \\ \hline
95\% Confidence Interval Lower Bound & -0.4859081               \\ \hline
95\% Confidence Interval Upper Bound & 0.2519416                \\ \hline
Mean of group with targets (x)       & 1.158098                 \\ \hline
Mean of group without targets (y)    & 1.275081                 \\ \hline
\end{tabular}
\caption{T-Test Results for Number of Victims}
\label{tab:ttest_results}
\end{table}

Differences between weapon types with respect to a the number of victims. Initially, the Kruskal-Wallis test \cite{kruskal1952use}, a non-parametric method, was used due to the non-normal distribution of our data across groups. The test revealed statistically significant differences among the groups. Next, Dunn's post-hoc tests \cite{dunn1964multiple} were performed to investigate which specific groups differed from each other. The following table presents the results of these post-hoc comparisons, with p-values adjusted using the Bonferroni\cite{dunn1964multiple} correction method to address the issue of multiple comparisons and control the error rate. Below are the findings that showed significance.

\begin{table}[H]
\centering
\begin{tabular}{lccc}
\hline
Comparison & Z value & P.unadj. & P.adj \\
\hline
Handgun - Other & 5.69 & $1.26 \times 10^{-8}$ & $3.53 \times 10^{-7}$ \\
Multiple Handguns - Other & 5.83 & $5.52 \times 10^{-9}$ & $1.54 \times 10^{-7}$ \\
Multiple Rifles - Other & 3.24 & $1.20 \times 10^{-3}$ & $3.36 \times 10^{-2}$ \\
Multple Unknown - Other & 4.94 & $7.68 \times 10^{-7}$ & $2.15 \times 10^{-5}$ \\
Other - Rifle & -3.17 & $1.52 \times 10^{-3}$ & $4.27 \times 10^{-2}$ \\
Other - Shotgun & -4.71 & $2.43 \times 10^{-6}$ & $6.80 \times 10^{-5}$ \\
Other - Various & -6.91 & $4.90 \times 10^{-12}$ & $1.37 \times 10^{-10}$ \\
Rifle - Various & -4.65 & $3.27 \times 10^{-6}$ & $9.16 \times 10^{-5}$ \\
\hline
\end{tabular}
\caption{Significant findings from Dunn's post-hoc test}
\label{tab:significant_findings}
\end{table}

\subsubsection{1966 - 2017 Data}
The process of linear regression was repeated for the data, but only through 2017, the rationale being that there was a stark uptick in school shooting events pre and post 2018. This approach could potentially discern differences in factors contributing to school shooting casualties. This process yielded the same factors for casualties.

\begin{table}[H]
\centering
\caption{ANOVA Table for Pre-2018 Events}
\label{anova-table}
\begin{tabular}{|l|c|c|c|c|c|}
\hline
\textbf{Variable} & \textbf{Df} & \textbf{Sum Sq} & \textbf{Mean Sq} & \textbf{F value} & \textbf{Pr(>F)} \\ 
\hline
Shooter\_Gender & 3 & 13.47 & 4.49 & 2.6289 & 0.049288 * \\ 
\hline
Weapon\_Type & 7 & 1639.10 & 234.16 & 137.0833 & < 2.2e-16 *** \\ 
\hline
Targets & 3 & 233.61 & 77.87 & 45.5870 & < 2.2e-16 *** \\ 
\hline
Accomplice & 2 & 6.72 & 3.36 & 1.9667 & 0.140725 \\ 
\hline
Bullied & 1 & 16.83 & 16.83 & 9.8530 & 0.001769 ** \\ 
\hline
Shots\_Fired & 1 & 594.01 & 594.01 & 347.7551 & < 2.2e-16 *** \\ 
\hline
Residuals & 676 & 1154.70 & 1.71 & & \\ 
\hline
\end{tabular}
\end{table}

\subsubsection{2018 - 2023 Data}
After repeating the process linear regression and iteratively removing factors of low significance, for the post-2018, data, there were changes in the factors that affected casualty count. The new factors included:
\begin{itemize}
    \item Shooter age
    \item Race
    \item Gang Related
\end{itemize}

It should also, be noted that shooter gender dropped off as a factor for this portion of the analysis.

\begin{table}[H]
\centering
\caption{ANOVA Table for Post-2018 Data}
\label{anovaLinPos2018}
\begin{tabular}{|l|c|c|c|c|c|}
\hline
\textbf{Variable} & \textbf{Df} & \textbf{Sum Sq} & \textbf{Mean Sq} & \textbf{F value} & \textbf{Pr(>F)} \\ 
\hline
Shooter\_Age & 1 & 12.03 & 12.03 & 6.9246 & 0.0106411 * \\ 
\hline
Weapon\_Type & 6 & 384.27 & 64.05 & 36.8806 & < 2.2e-16 *** \\ 
\hline
Race & 4 & 45.29 & 11.32 & 6.5197 & 0.0001825 *** \\ 
\hline
During\_Classes & 1 & 4.22 & 4.22 & 2.4320 & 0.1238143 \\ 
\hline
Targets & 3 & 78.36 & 26.12 & 15.0417 & 1.628e-07 *** \\ 
\hline
Accomplice & 1 & 2.97 & 2.97 & 1.7093 & 0.1957522 \\ 
\hline
Gang\_Related & 1 & 4.01 & 4.01 & 2.3094 & 0.1335164 \\ 
\hline
Shots\_Fired & 1 & 556.41 & 556.41 & 320.4094 & < 2.2e-16 *** \\ 
\hline
Residuals & 64 & 111.14 & 1.74 & & \\ 
\hline
\end{tabular}
\end{table}

\subsection{Negative Binomial Regression}

\subsubsection{All Events}
For the negative binomial regression model, the number of casualties had to be converted to a binary variable which would reflect whether or not there was a casualty in an event. After removing statistically insignificant variables the model was left with two factors that contribute to the presence of casualties.
\begin{itemize}
    \item Shooter Age
    \item Targeted Attacks
\end{itemize}

\begin{table}[H]
\centering
\caption{ANOVA Table of All Events}
\begin{tabular}{|l|r|r|r|r|r|}
\hline
\textbf{Variable} & \textbf{Df} & \textbf{Deviance} & \textbf{Resid. Df} & \textbf{Resid. Dev} & \textbf{Pr(>Chi)} \\ \hline
NULL              &             &                   & 2173               & 1198.0              &                   \\ \hline
Shooter\_Age      & 1           & 0.257             & 2172               & 1197.8              & 0.6124            \\ \hline
Targets           & 3           & 173.326           & 2169               & 1024.4              & <2e-16        \\ \hline
\end{tabular}

\label{tab:anovaAllNB}
\end{table}

\paragraph{1966 - 2017 Data}
For the negative binomial regression of the pre-2018 data, the targeted victims factor remains, but the shooter age is replaced by the shooter being killed.
\begin{table}[H]
\centering
\caption{ANOVA Table For Pre-2018 Events}
\begin{tabular}{|l|r|r|r|r|r|}
\hline
\textbf{Variable}    & \textbf{Df} & \textbf{Deviance} & \textbf{Resid. Df} & \textbf{Resid. Dev} & \textbf{Pr(>Chi)}   \\ \hline
NULL                 &             &                   & 1211               & 508.64              &                     \\ \hline
Shooter\_Killed      & 1           & 44.491            & 1210               & 464.15              & 2.556e-11       \\ \hline
Targets              & 3           & 40.835            & 1207               & 423.32              & 7.089e-09        \\ \hline
\end{tabular}

\label{tab:anovaPre2018NB}
\end{table}

\subsubsection{2018 - 2023 Data}
For the negative binomial regression of the post-2018 data, targeted attacks remain a statistically significant factor, but race is now added in place of the shooter being killed as a factor. 

\begin{table}[H]
\centering
\caption{ANOVA Table For Post-2018 Data}
\begin{tabular}{|l|r|r|r|r|r|}
\hline
\textbf{Variable} & \textbf{Df} & \textbf{Deviance} & \textbf{Resid. Df} & \textbf{Resid. Dev} & \textbf{Pr(>Chi)}   \\ \hline
NULL              &             &                   & 200                & 130.149             &                     \\ \hline
Race              & 4           & 12.954            & 196                & 117.196             & 0.01151 *           \\ \hline
Targets           & 3           & 31.437            & 193                & 85.759              & 6.878e-07 ***       \\ \hline
\end{tabular}
\label{tab:anova}
\end{table}

\section{Discussion}

The K12 School Shooting database is a robust dataset with the potential for study from a variety of angles. While attempting multiple data modeling techniques and statistical analysis tools with RStudio, the relationships between many combinations of the available variables can investigated. The extent of this study serves as a preliminary demonstration of one of the possible directions that can be taken while trying to gain a foothold in understanding this sensitive topic. While correlation is not the same as causation it can be useful to use these seemingly related factors as a jumping off point. The investigation into the data set also provides insight into common misconceptions of who commits these crimes and how these events unfold.

\subsection{Preliminary Data Analysis}

\subsubsection{Regions Prone to School Shootings}The geographical heat maps\ref{fig:AllHeatMap}\ref{fig:heatmapsoldandnew} highlight a fair amount of consistency through time for the school shooting events with the only major differences observed in the region in the state of Colorado for events from 2018 - 2013. The reasons for this difference are not obvious are beyond the scope of this study. Future studies could also explore the prevalence of events based on weapons of choice, firearms laws, demographic shifts, and regional trends.

\subsubsection{Increase in Annual Occurrences of School Shootings} Linear and exponential modeling of events per year was performed in RStudio. The exponential model provided the best fit, showing a significant increase in events starting from roughly 2018. The equation for exponential growth ( $y=22.49+0.00001855e^{0.2872t}$) suggests that the year 2024 will see about 341 school shooting events. To date (11-April 2014) there have been 88\cite{Database}, so unfortunately these findings are on track with the current state of affairs.

\subsubsection{Time of Year Trends in School Shooting Events}
An analysis of incidents per month revealed July as the month with the lowest occurrences (70) and September as the highest (344). Summary statistics include a mean of 214.67, a median of 217, and a standard deviation of 79.59. A $\chi^{2}$ test suggested seasonal trends, with January, May, September, and October experiencing higher rates of shootings. This insight suggests increased vigilance for student, faculty, and staff safety during "peak" periods of time when school shooting events appear statistically more likely to occur. 

\subsubsection{Casualty Information}
In the school shooting event casualty category, the total number of people killed was 780, with 2152 wounded. Statistical data showed a maximum of 26 killed in a single incident and a maximum of 74 wounded (separate events). Demographic analysis revealed that the majority of victims were male (75.94\%), students (66.30\%), and teens (54.02\%), with black victims constituting 50.29\%, which appears to be a growing trend\cite{katsiyannis2023examination}. 

\subsubsection{Shooter Demographics}
Shooters were predominantly in their late teens with a mean age of 20.8 years and a median of 17 years. The majority were male, and among the known racial data, 50.55\% were black. Most shooters were students (1106), with a significant number having no relation to the school (575). Background information indicated a low incidence of bullying, gang-related activity, and domestic violence among shooters. This turns the narrative of the bullied, troubled teen or gang member with a history of violence on its head and demonstrates that there is no single face of a school shooter.

\subsubsection{Shooting Event Details and Outcomes}
In analysis of the details around the shooting events it was found that 61.63\% of shootings occurred during class sessions, with only 22.41\% involving an accomplice. This shows that shooters tend to act alone and also choose to engage in a school shooting when classes are in session and there is access to more people on campus. Hostages were rarely taken (2.36\%) which further drills down upon potential motivating factors. The shooter was killed in only 9.86\% of events which could indicate how law enforcement and other entities handle and neutralize these events. Handguns were the primary weapon choice (72.65\%). This make logical sense in that it is easier to sneak a smaller firearm onto a campus. Over half of the incidents (56.77\%) involved targeted victims and yet the data analysis (discussed below) shows that this does not seem to directly lead to increased casualties which implies that the shooter is not always successful in their attempts on their targets. The majority of shootings (60.67\%) occurred outside but on school grounds. Entering a building presents more of potential challenge to a shooter as well as further opportunities to get apprehended.

\subsection{Regression Models}
One common factor that appears to be consistent between all of the models to this point is whether or not the shooter(s) had specific targets. This aligns with the reality of a shooter entering an event having a specific goal of hurting specific individuals. This finding aligns with previous meta analysis of school/active shooter events \cite{bonanno2014school}. Which would lead to a higher likelihood of either more casualties as shown in the linear regression, or, in the case of the negative binomial regression, any casualties at all. However, when further analyzed with a Welch's t-test,  \ref{tab:ttest_results} this finding did not hold up. There are multiple possible explanations for why which would be useful in further study.

The slight changes in significant variables pre and post 2018 are also noteworthy. With the addition of shooter's race, gang affiliation, and dropping off of bullying \ref{anovaLinPos2018} . The shooter's ages and the presence of an accomplice also a common thread that is noticed in the linear regression models. 

For the negative binomial regression, it is noted that the amount of statistically significant factors is down to as little as two, of which, targeted attacks remain a factor. It is also important to note that further study should attempt a different model due to the goodness of fit of the negative binomial being sub-optimal. Most of the coefficients for the model were near zero. 

There are still many areas that remain for identifying and refining the best modeling technique for this dataset. Further study should involve different approaches to coding the data and enhanced scrutiny of co-linearity  of the variables.  There are also adaptations that can be made from sports analytics such as least absolute errors\cite{bassett1997robust}.

\section{Conclusion}
School shootings are a unsettling and sensitive yet important topic for understanding. With a wealth of data available which detail the events, much can be learned that could lead to policy changes, outreach programs for those at risk, and increased overall understand of this growing and troubling phenomenon. The data analysis performed in this study is merely a start, but the image of a school shooter often portrayed in the media does not match what the data shows. This finding alone leads to a helpful clarification of the contributing factors. Further study into these factors could lead to a scoring systems that takes the various characteristics of an individual believed to be at risk and could assign true risk factors. Furthermore, continued refining of the data modeling could lead to better campus layouts and student safety policies geared toward protecting and minimizing harm when events do take place. School shootings are not going away in the foreseeable future. It is important to make the most of resources and policies to help lessen the damage they can cause.

%Bibliography
\bibliographystyle{unsrt}  
\bibliography{references}

\end{document}